\documentclass{jps-cp}
\usepackage{txfonts}

\title{Associated Quarkonium Hadroproduction at High-Energy Colliders
}

\author{Nodoka \textsc{Yamanaka}$^{1}$,
Jean-Philippe \textsc{Lansberg}$^{1}$,
Hua-Sheng \textsc{Shao}$^{2}$, and
Yu-Jie \textsc{Zhang}$^{3}$
}

\inst{
$^{1}$IPNO, CNRS-IN2P3, Univ. Paris-Sud, Universit\'e Paris-Saclay, 91406 Orsay Cedex, France\\
$^{2}$Laboratoire de Physique Th\'eorique et Hautes Energies (LPTHE), UMR 7589, Sorbonne Universit\'e et CNRS, 4 place Jussieu, 75252 Paris Cedex 05, France\\
$^{3}$Beijing Key Laboratory of Advanced Nuclear Energy Materials and Physics, and School of Physics, Beihang University, Beijing 100191, China
}

\email{
yamanaka@ipno.in2p3.fr
}

\recdate{April 27, 2019}

\abst{
Quarkonium production in proton-proton collision is interesting in profiling the partons inside the nucleon.
Recently, the impact of double parton scatterings (DPSs) was suggested by experimental data of 
associated quarkonium production ($J/\psi + Z$, $ J/\psi + W$, and $J/\psi + J/\psi$) at the LHC and Tevatron, in addition to single parton scatterings (SPSs).
In this proceedings contribution, we review the extraction of the effective parameter of the DPS ($\sigma_{\rm eff}$) through the evaluation of the SPS contributions under quark-hadron duality.
}

\kword{Nucleon structure, double parton scattering, quarkonia}

\begin{document}
\maketitle

\voffset=-20mm

\section{Introduction}

The prime motivation to study quarkonium production is to unveil novel nonperturbative and perturbative features of QCD \cite{Lansberg:2019adr,Andronic:2015wma,Brambilla:2010cs,Lansberg:2006dh}.
In this context, an interesting class of processes is that of the associated quarkonium production, which is being studied to probe double parton scatterings (DPS) \cite{Diehl:2014vaa,Diehl:2017wew,Buffing:2017mqm} and even triple parton scatterings \cite{Shao:2019qob}.
A representative case is di-$J/\psi$ production, which was measured in many experiments (NA3 \cite{Badier:1985ri}, D0 \cite{Abazov:2014qba}, CMS \cite{Khachatryan:2014iia}, ATLAS \cite{Aaboud:2016fzt}, and LHCb \cite{Aaij:2011yc,Aaij:2016bqq}), and was studied in many theoretical works \cite{Kom:2011bd,Baranov:2015cle,Borschensky:2016nkv,Lansberg:2014swa}.
Recently, experimental data of associated production with vector bosons were released by the ATLAS Collaboration ($J/\psi +W$ \cite{Aad:2014rua} and $J/\psi +Z$ \cite{Aad:2014kba}).
The single parton scattering (SPS) contributions to these processes were theoretically computed in NRQCD \cite{Li:2010hc,Mao:2011kf,Gong:2012ah,Lansberg:2013wva,Lansberg:2016muq,Shao:2016wor}, and the predictions have difficulties in explaining the yields in several regions of the phase space.
This proceedings contribution summarizes the results of the calculations of the SPS of $J/\psi +W$, $J/\psi +Z$, and $J/\psi +J/\psi$ \cite{Lansberg:2014swa,Lansberg:2013qka,He:2015qya,Sun:2014gca} production in the color evaporation model (CEM) which provides us indirect informations about the DPS.

\section{The double parton scattering}

Let us parametrize the DPS.
If we assume two uncorrelated parton scatterings, the DPS cross section can be written as 
\begin{equation}
\sigma_{\rm DPS} (A+B) = \frac{1}{1+\delta_{AB}} \frac{\sigma (A) \sigma (B)}{\sigma_{\rm eff}}
,
\end{equation}
with $\delta_{AB} =1$ for the case where we have $A=B$ in the final state, where $A$ or $B$ (or both) is a quarkonium.

\section{The color evaporation model}

The CEM is a model to calculate heavy quarkonium production processes based on quark-hadron duality \cite{Lansberg:2006dh,Fritzsch:1977ay,Halzen:1977rs,Amundson:1995em,Amundson:1996qr}.
In this model, the quarkonium ${\cal Q}$ is produced as a quark-antiquark pair $Q\bar Q$ having its invariant mass below the open-heavy flavor threshold $2 m_{\rm thr.}$.
The cross section in the model is given by
\begin{equation}
\sigma^{\rm (N)LO,\ \frac{direct}{prompt}}_{\cal Q} = {\cal P}^{\rm (N)LO,\frac{direct}{prompt}}_{\cal Q}\int_{2m_Q}^{2m_{\rm thr.}} \frac{d\sigma_{Q\bar Q}^{\rm (N)LO}}{d m_{Q\bar Q}}d m_{Q\bar Q}
,
\end{equation}
where we assume universal parameters  ${\cal P}^{\rm (N)LO,{prompt}}_{\cal Q}$.
For $J/\psi$, we have ${\cal P}^{\rm (N)LO,{prompt}}_{J/\psi} = 0.014$ (LO), 0.009 (NLO) \cite{Lansberg:2016rcx}, obtained from the fit of the single inclusive $J/\psi$ hadroproduction data.
A caveat is that the single-quarkonium production cross section predicted by the model overshoots the experimental data at high transverse momentum $p_T$ \cite{Lansberg:2006dh,Andronic:2015wma,Lansberg:2016rcx}.
It is understood that the dominance of the gluon fragmentation in the model yields too hard a $p_T$ spectrum, which should also apply to the associated quarkonium production with vector bosons, discussed in the next section.

\section{Analysis of the ATLAS data for $J/\psi +Z$ and $J/\psi +W$ productions in the CEM}

Let us now consider the $J/\psi +Z$ and $J/\psi +W$ productions.
As we mentioned in the previous section, the single quarkonium production in the CEM is dominated by the gluon fragmentation topologies at large $p_T$, which also happens for the cases of $J/\psi +Z$ and $J/\psi +W$.
Since the CEM predictions overshoot the experimental data at high $p_T$, we can set conservative upper limits to the SPS contribution of both these processes.
The SPS is evaluated at NLO in $\alpha_s$ with {\small \sc MadGraph5\_aMC@NLO} \cite{Alwall:2014hca}.

\begin{table}[tbh]
\begin{center}
\caption{
Results of the NLO calculations of the $J/\psi +Z$ and $J/\psi +W$ production cross sections in the CEM.
The experimental data of ATLAS are also shown for comparison with the combined statistical and systematic uncertainties.
}
\label{table:comparison}
\begin{tabular}{l|ll}
\hline
& ATLAS & NLO CEM \\
\hline
$J/\psi +Z$ & $1.6 \pm 0.4$ pb \cite{Aad:2014kba} & $0.19^{+0.05}_{-0.04}$ pb \cite{Lansberg:2016rcx}\\
$J/\psi +W$ & $4.5 ^{+1.9}_{-1.5}$ pb \cite{Aad:2014rua} & $0.28 \pm 0.07$ pb \cite{Lansberg:2017chq}\\
\hline
\end{tabular}
\end{center}
\end{table}

\begin{figure}[htb]
\begin{center}
\includegraphics[clip,width=.45\columnwidth]{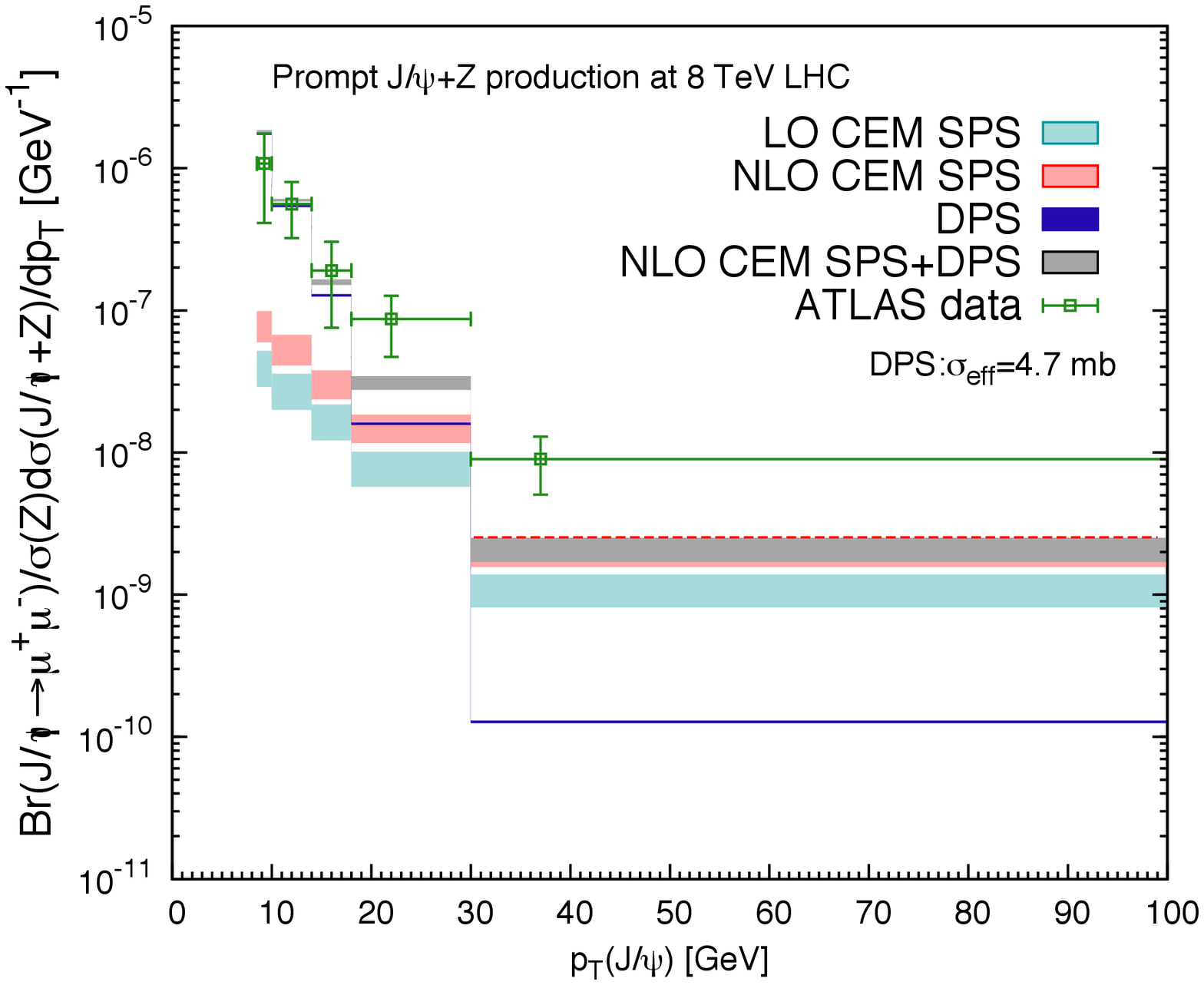}
\includegraphics[clip,width=.45\columnwidth]{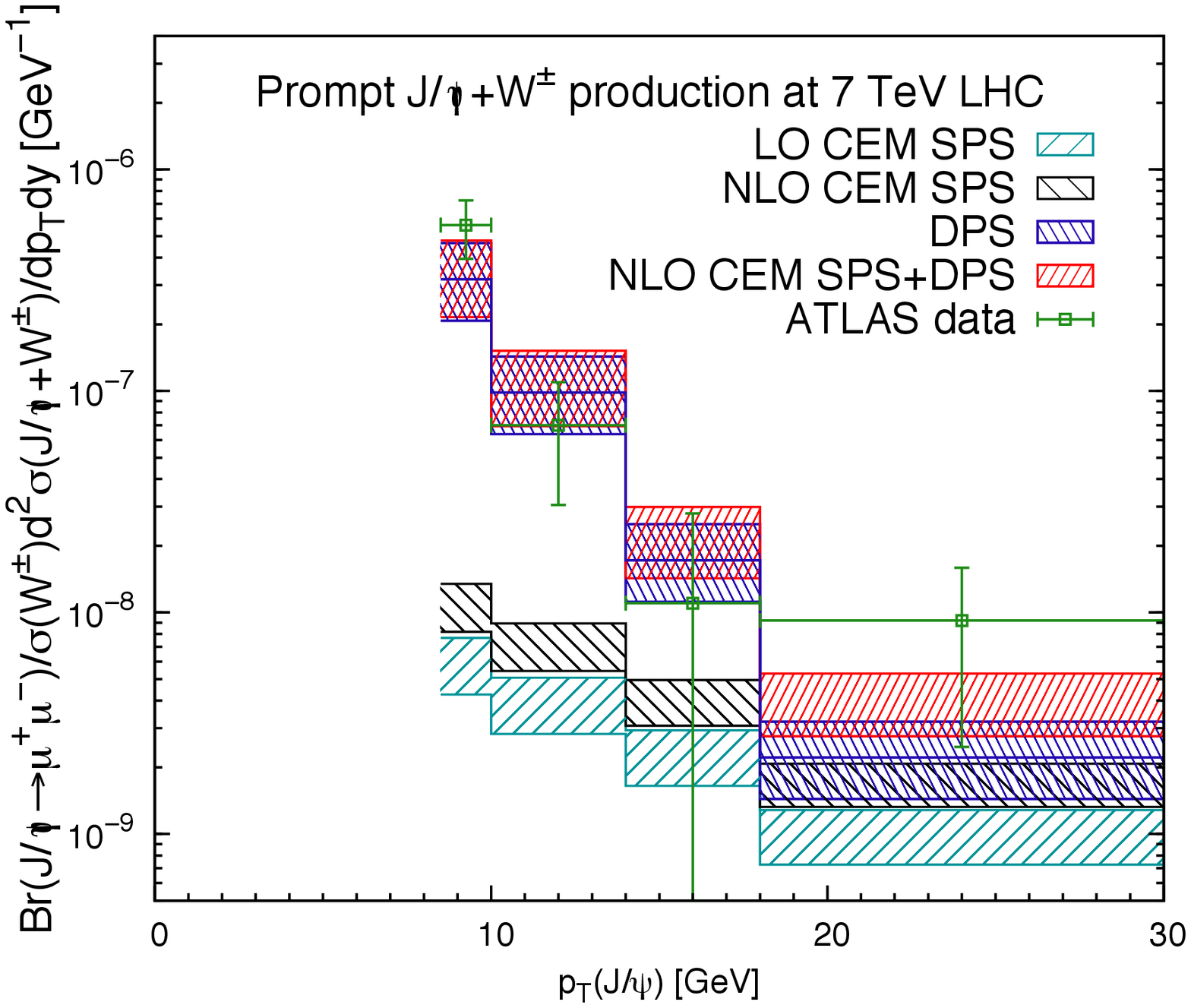}
\caption{
The $p_T$ distribution of the $J/\psi$ in the $J/\psi +Z$ \cite{Lansberg:2016rcx} (left panel) and $J/\psi +W$ \cite{Lansberg:2017chq} (right panel) production cross section in the CEM.
The ATLAS experimental data \cite{Aad:2014kba,Aad:2014rua} are also displayed for comparison.
}
\label{fig:pT_distribution}
\end{center}
\end{figure}

Table \ref{table:comparison} shows the results of the associated $J/\psi$ productions with vector bosons.
We see that the NLO CEM SPS predictions alone are smaller than the ATLAS experimental data (see also Fig. \ref{fig:pT_distribution}).

Let us now fit $\sigma_{\rm eff}$ by assuming that the DPS fills the gap between the SPS and the measured total cross section.
The result is shown in Fig. \ref{fig:pT_distribution}.
We obtain $\sigma_{\rm eff} = (4.7 ^{+2.4}_{-1.5})$ mb \cite{Lansberg:2016rcx} (${J/\psi +Z}$) and $\sigma_{\rm eff} = (6.1 ^{+3.3}_{-1.9})$ mb \cite{Lansberg:2017chq} (${J/\psi +W}$).

\section{Analysis of di-$ J/\psi$ production in the CEM}

Let us now evaluate the di-$J/\psi$ production in the CEM.
The regions of the phase space of interest are at the large invariant mass $M_{\psi \psi}$ and rapidity separation $\Delta y$, where the experimental data of CMS and ATLAS are overshooting the color singlet model SPS prediction \cite{Khachatryan:2014iia,Lansberg:2013qka,Lansberg:2014swa,Aaboud:2016fzt}.

\begin{figure}[htb]
\vspace{-2em}
\begin{center}
\includegraphics[clip,width=.45\columnwidth]{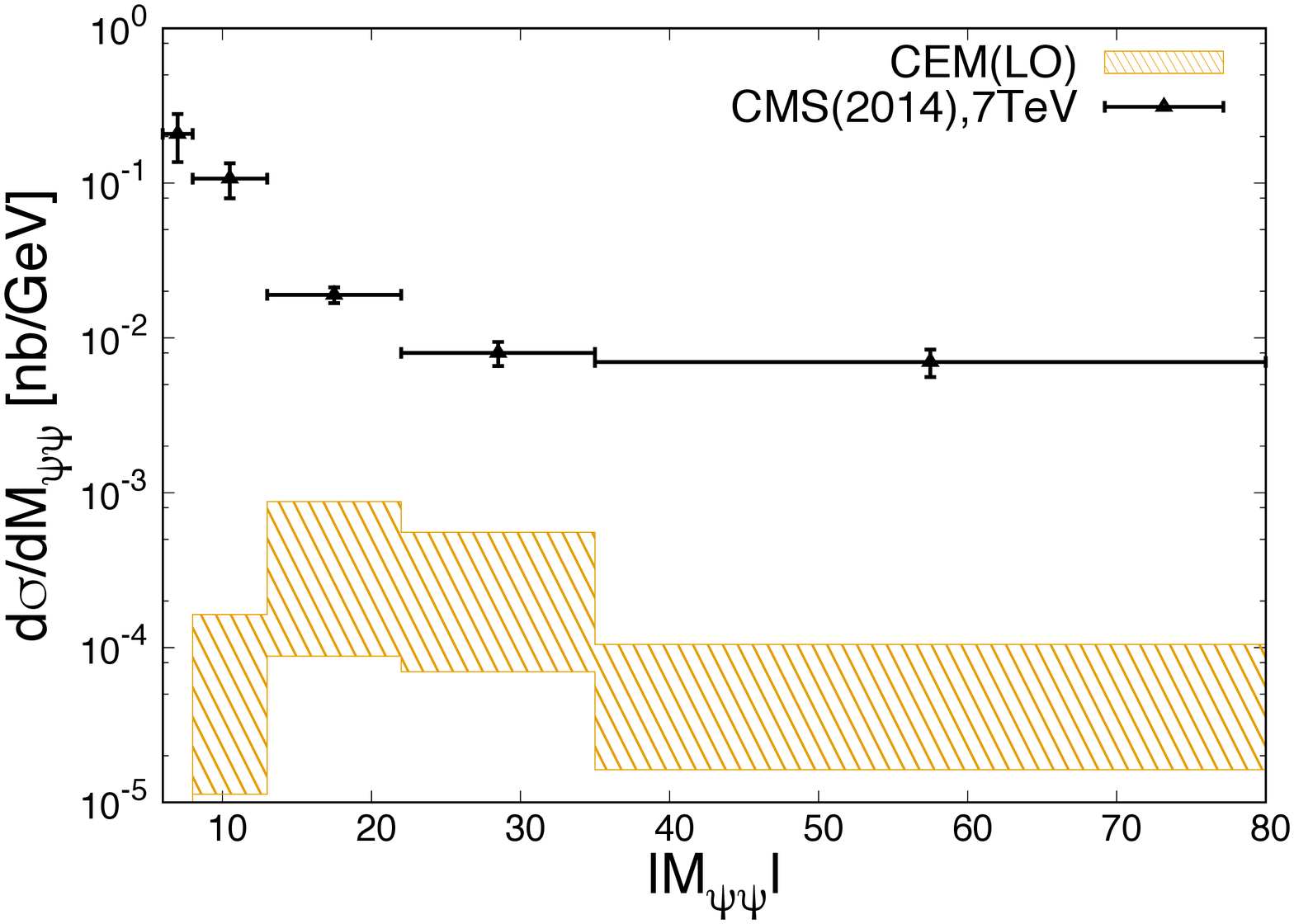}
\includegraphics[clip,width=.45\columnwidth]{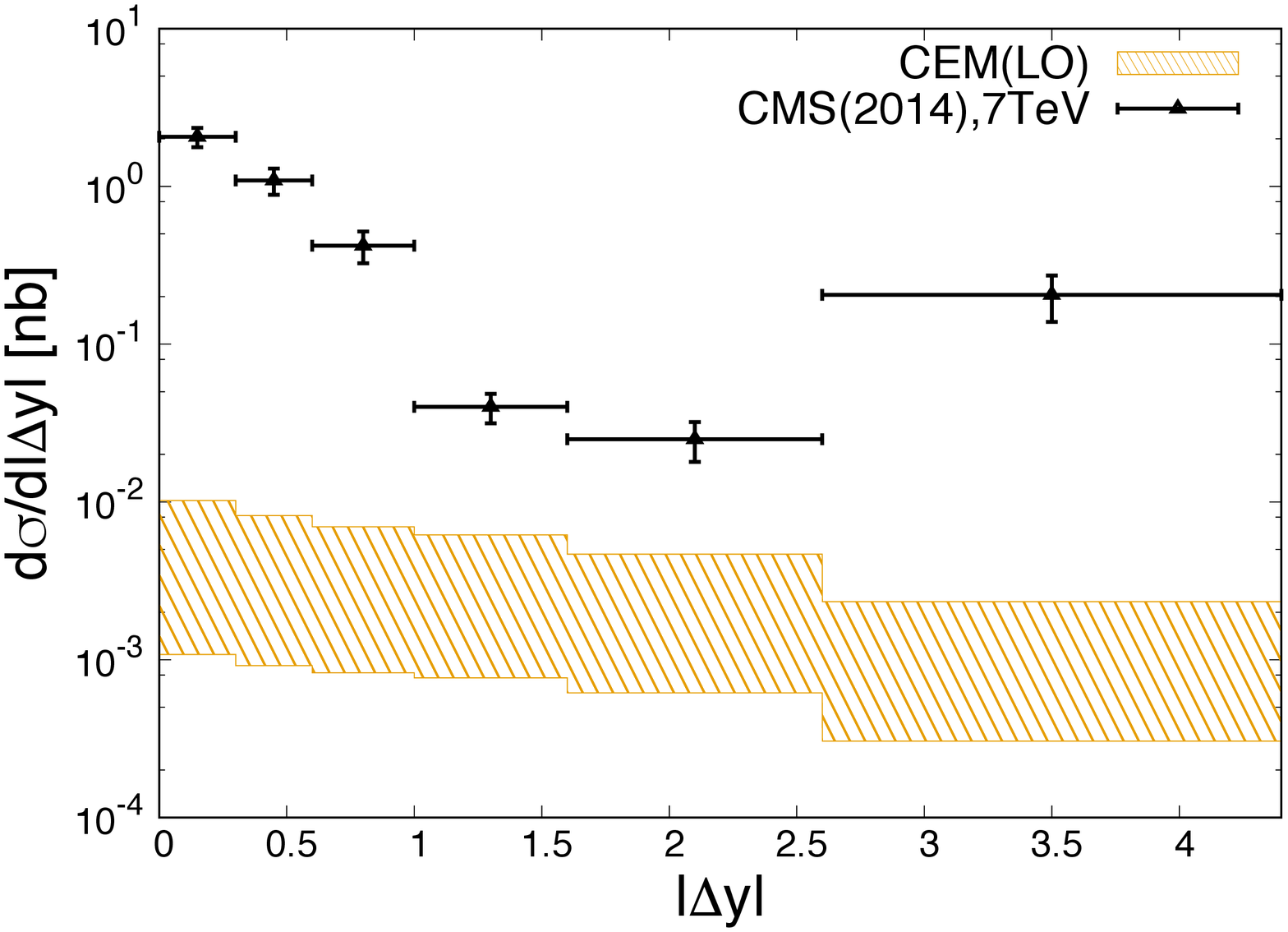}
\vspace{-1.5em}
\caption{
The invariant mass (left panel) and $\Delta y$ (right panel) differential cross sections of di-$J/\psi$ production (CMS setup, $\sqrt{s} = 7$ TeV).
}
\label{fig:double_Jpsi_CMS7TeV}
\end{center}
\vspace{-3em}
\end{figure}

\begin{figure}[tbh]
\vspace{-1em}
\begin{center}
\includegraphics[clip,width=.5\columnwidth]{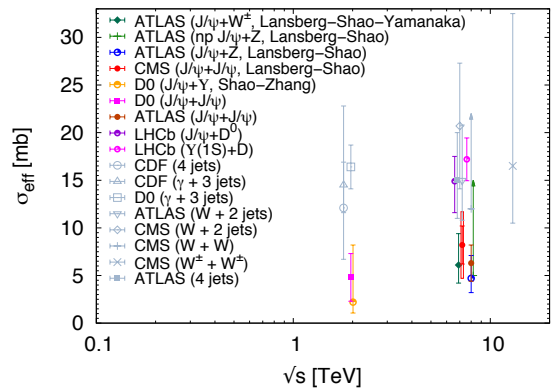}
\caption{Summary of several extractions of $\sigma_{\rm eff}$. 
Quarkonium related extractions are shown in color (see Ref. \cite{Lansberg:2017chq}).
}
\label{fig:sigmaeff_summary}
\end{center}
\end{figure}

By computing the SPS contribution to the di-$J/\psi$ production at LO, we obtain the result of Fig.~\ref{fig:double_Jpsi_CMS7TeV}.
No particular enhancements at large $M_{\psi \psi}$ and $\Delta y$ are seen in the CEM.
Our result is suggesting the dominance of the DPS in these regions of the di-$J/\psi$ production.
By assuming the dominance of the DPS, the $\sigma_{\rm eff}$ value extracted from the CMS \cite{Khachatryan:2014iia} ($\sigma_{\rm eff} = (8.2 \pm 2.0_{\rm stat} \pm 2.9_{\rm sys})$ mb \cite{Lansberg:2014swa}), D0 ($\sigma_{\rm eff} = (4.8 \pm 0.5_{\rm stat} \pm 2.5_{\rm sys})$ mb) \cite{Abazov:2014qba}, and ATLAS Collaborations ($\sigma_{\rm eff} = (6.3 \pm 1.6_{\rm stat} \pm 1.0_{\rm sys})$ mb) \cite{Aaboud:2016fzt} are all consistent with each other, as well as with those of the $J/\psi +W$ and $J/\psi +Z$ productions.
In Fig. \ref{fig:sigmaeff_summary}, we summarize the extractions of $\sigma_{\rm eff}$ from different processes and experimental data.

\section{Conclusion}

To summarize, we analyzed the production processes of $J/\psi + W/Z$ (NLO) and $J/\psi + J/\psi$ (LO) in the CEM. 
For the case of $J/\psi + W/Z$, it is possible to extract the DPS yield from the experimental data by setting an upper limit on the SPS contribution.
We obtained 
$\sigma_{\rm eff} = (4.7 ^{+2.4}_{-1.5})$ mb (${J/\psi +Z}$), and $\sigma_{\rm eff} = (6.1 ^{+3.3}_{-1.9})$ mb (${J/\psi +W}$), which emphasizes the importance of the DPS and is compatible with other extractions from other central rapidity quarkonium data.
This $\sigma_{\rm eff}$ is also in agreement with the enhancement of the di-$J/\psi$ production at large $\Delta y$ and invariant mass.

\end{document}